\documentstyle[12pt]{article}

  \def\half{{1 \over 2}}
  \def\i3{{1 \over 3}}
  \def\sqr2{\sqrt{2}}
  \def\isqr2{{1 \over \sqrt{2}}}

  \def\bv{{\mbox {\boldmath $v$}}}
  \def\tO{\tilde{\Omega}}
  \def\o{\omega}
  \def\th{\theta}
  \def\to{\tilde{\omega}}
  \def\tv{\tilde{\mbox {\boldmath $v$}}}

\begin{document}

\vspace*{0.7cm}
\begin{Large}
\begin{center}
Approximate Sum Rules of CKM Matrix Elements\\
from Quasi-Democratic Mass Matrices
\end{center}
\end{Large}

\vskip 0.8cm
\begin{large}
\begin{center}
Ikuo S. Sogami\footnote{E-mail address: sogami@cc.kyoto-su.ac.jp},
Kouzou Nishida,
Hajime Tanaka\footnote{E-mail address: gen@cc.kyoto-su.ac.jp} \\
and
Tadatomi Shinohara\footnote{E-mail address: sinohara@cc.kyoto-su.ac.jp} \\
\vskip 0.3cm
{\it Department of Physics, Kyoto Sangyo University, Kyoto 603}
\end{center}
\end{large}

\vskip 1.5cm
\begin{abstract}
To extract sum rules of CKM matrix elements, eigenvalue problems for
quasi-democratic mass matrices are solved in the first order perturbation
approximation with respect to small deviations from the democratic limit.
Mass spectra of up and down quark sectors and the CKM matrix are shown to have
clear and distinctive hierarchical structures. Numerical analysis shows that
the absolute values of calculated CKM matrix elements fit the experimental
data quite well. The order of the magnitude of the Jarlskog parameter is
estimated by the relation 
$|J| \approx \sqrt{2}(m_c/m_t + m_s/m_b)|V_{us}|^2|V_{cb}|/4$.
\end{abstract}

\newpage

\section{Introduction}

\qquad
Fundamental fermions, quarks and leptons, exist with a broad mass
spectrum, ranging from zero or almost zero masses of neutrinos to 180GeV
of the top quark. Magnitude of the weak Cabbibo-Kobayashi-Maskawa
(CKM) matrix elements~\cite{rf:1,rf:2} tends to decrease rapidly
from the diagonal to the off-diagonal direction. It is impossible to explain
such hierarchical structures of fundamental fermions within the framework
of the standard model. To do so it is necessary to postulate
some working hypothesis on the mass matrices from outside of the model.\par

Among various forms of mass matrices~{\cite{rf:3}$\sim$\cite{rf:6}}, democratic
mass matrices with small correction terms explain such hierarchical
structures in a simple and systematic way~{\cite{rf:7}$\sim$\cite{rf:19}}.
The calculability of the CKM matrix elements in terms of quark mass ratios is
examined under the hypothesis of the universal strength of Yukawa couplings
in which all Yukawa coupling constants are assumed to have equal
moduli~{\cite{rf:15}$\sim$\cite{rf:18}}.  It was shown that it is possible
to find the CKM matrix elements from quasi-democratic mass matrices which are
consistent with present experimental data to high accuracy~\cite{rf:19}.
In this article, sum rules of the CKM matrix elements are extracted
from Hermitian quasi-democratic mass matrices which have small deviation terms
and phases in off-diagonal elements. For this purpose we solve mass eigenvalue
problems in a first order perturbation approximation with respect to
small deviations around the democratic limit.\par

We postulate here that mass matrices for up and down quark sectors are
quasi-democratic and represented generically in the forms
\begin{equation}
    {\cal M}_q = M_q\tO_q, \quad (q = u, d)
\end{equation}
where $M_q$ is a mass scale for the $q$-sector and $\tO_q$ is the Hermitian
matrix
\begin{equation}
  \tO_q =\i3
         \left(
         \begin{array}{ccc}
          1  &  a_3^q e^{i\delta_{12}^q}
             & a_2^q e^{-i\delta_{31}^q}  \\
         \noalign{\vskip 0.2cm}
          a_3^q e^{-i\delta_{12}^q} & 1
             & a_1^q e^{i\delta_{23}^q} \\
         \noalign{\vskip 0.2cm}
          a_2^q e^{i\delta_{31}^q}
             & a_1^q e^{-i\delta_{23}^q} & 1
          \end{array}
         \right)
   \label{quasi-democratic}
\end{equation}
with phases satisfying the restriction
\begin{equation}
  \delta_{12}^q + \delta_{23}^q + \delta_{31}^q = 0.
  \label{phase}
\end{equation}
The real parameters $a_j^q$ and $\delta_{jk}^q$ are presumed to take values
which deviate slightly from the democratic limit $a_j^q = 1$ and
$\delta_{jk}^q = 0$.
\par

Note that, without loss of generality, rephrasing of chiral quark fields
enables us to reduce the phases $\delta_{jk}^q$ as
\begin{equation}
   \delta_{12}^q = 0,\ \ 
   \delta_{23}^q = \phi_q,\ \ 
   \delta_{31}^q = - \phi_q
\end{equation}
by a single real number $\phi_q$ for each $q$-sector. This type of mass
matrices was investigated first by Branco, Silva and Rebelo~\cite{rf:15} for
the quark masses squared in their study of the universal strength for Yukawa
coupling cosntants. Then Teshima and Sakai~\cite{rf:19} applied it to quark
masses and carried out numerical analysis. \par

In the first order perturbation approximation, we obtain CKM matrix
elements parametrized by quantities which naturally describe small deviations
around the democratic limit. We obtain mass spectra of up and down quark
sectors and the CKM matrix. These spectra have distinctive hierarchical
structures. It is possible to express the absolute values of the CKM matrix
elements in terms of three parameters. Consequently we are able to derive
six independent sum rules for the absolute values of nine matrix elements.
All of these agree well with available experimental data.
In \S 2 we formulate and solve the mass eigenvalue problems for
quasi-democratic quark mass matrices.
In \S 3, we obtain the CKM matrix and extract sum rules from it.
Comparison of our results with the experimental data is made in \S 4,
and discussion is given in \S 5. \par

\section{Mass eigenvalue problems}

\qquad
It is possible to obtain exact solutions of the eigenvalue problem for the
matrix $\tO_q$~\cite{rf:18}. However, the exact eigenvectors take forms which
are too complicated to be used to extract simple and physically-meaningful
analytical relations among the CKM matrix elements. Hence we attempt here to
solve the eigenvalue problems in a first order perturbation approximation
with respect to small parameters $\lambda_q \propto a_1^q - a_2^q$. For this
purpose it turns out to be convenient to parameterize $a_j^q$ as
\begin{equation}
       a_1^q = {3 \over 2\sqr2}\rho_q\sin\th_q + 3\lambda_q,\ \ 
       a_2^q = {3 \over 2\sqr2}\rho_q\sin\th_q - 3\lambda_q,\ \ 
       a_3^q = 3\rho_q\cos\theta_q
\end{equation}
and then to decompose the matrix $\tO_q$ into unperturbative and perturbative
parts as
\begin{equation}
  \tO_q = \Omega_q + \Lambda_q,
\end{equation}
where
\begin{equation}
  \Omega_q = \i3
     \left(
       \begin{array}{ccc}
        1  & 3\rho_q\cos\th_q
           & \displaystyle{{3 \over 2\sqr2}}\rho_q\sin\th_q e^{i\phi_q}\\
       \noalign{\vskip 0.15cm}
        3\rho_q\cos\th_q  & 1 
           & \displaystyle{{3 \over 2\sqr2}}\rho_q\sin\th_q e^{i\phi_q}\\
       \noalign{\vskip 0.15cm}
        \displaystyle{{3 \over 2\sqr2}}\rho_q\sin\th_q e^{-i\phi_q}
           & \displaystyle{{3 \over 2\sqr2}}\rho_q\sin\th_q e^{-i\phi_q} & 1
       \end{array}
     \right)
\end{equation}
and
\begin{equation}
  \Lambda_q = \lambda_q
         \left(
          \begin{array}{ccc}
                  0        &        0       &  - e^{i\phi_q}\\
                \noalign{\vskip 0.1cm}
                  0        &        0       &    e^{i\phi_q}\\
                \noalign{\vskip 0.1cm}
            - e^{-i\phi_q} &   e^{-i\phi_q} &       0
          \end{array}
         \right).
\end{equation}
In this parametrization the democratic limit is realized when
\begin{equation}
  \rho_q = 1,\quad \cos\th_q = \i3,\quad \lambda_q = 0,\quad \phi_q = 0.
\end{equation}

Under the assumption $\lambda_q \ll 1$, which is confirmed later to be valid
for both up and down quark sectors, we solve the eigenvalue problems
\begin{equation}
    \tO_q\tv_j^q = \to_j^q\tv_j^q \qquad (q = u, d)
\end{equation}
in a perturbative approximation. The unperturbed eigenvalue problems
\begin{equation}
    \Omega_q\bv_j^q = \o_j^q\bv_j^q \qquad (q = u, d)
\end{equation}
are readily solved with the eigenvalues $\o_j^q $ and eigenvectors $\bv_j^q$
in the forms
\begin{equation}
 \begin{array}{ll}
  \o_1^q = \displaystyle{{1 \over 3}} - \rho_q\cos\th_q   & :\ \ 
  \bv_1^q= \displaystyle{{1 \over \sqr2}}
            \left( 
              \begin{array}{c}
                1  \\ 
                -1 \\ 
                0
              \end{array}
            \right) ,\\
  \noalign{\vskip 0.3cm}
  \o_2^q = \displaystyle{{1 \over 3}}
         + \displaystyle{{1 \over 2}}\rho_q(-1+\cos\th_q)   & :\ \ 
  \bv_2^q= \displaystyle{{1 \over \sqr2}}
            \left( 
              \begin{array}{c}
                \sin\displaystyle{\th_q \over 2}\\ 
                \noalign{\vskip 0.15cm}
                \sin\displaystyle{\th_q \over 2}\\ 
                \noalign{\vskip 0.15cm}
                -\sqr2 e^{-i\phi_q}\cos\displaystyle{\th_q \over 2}
              \end{array}
            \right) ,\\
  \noalign{\vskip 0.3cm}
  \o_3^q = \displaystyle{1 \over 3}
         + \displaystyle{1 \over 2}\rho_q(1+\cos\th_q)   & :\ \ 
  \bv_3^q= \displaystyle{1 \over \sqr2}
            \left( 
              \begin{array}{c}
                \cos\displaystyle{\th_q \over 2}\\ 
                \noalign{\vskip 0.15cm}
                \cos\displaystyle{\th_q \over 2}\\ 
                \noalign{\vskip 0.15cm}
                \sqr2 e^{-i\phi_q}\sin\displaystyle{\th_q \over 2}
              \end{array}
            \right) .
 \end{array}
\label{noneigen}
\end{equation}

In the first order perturbation with respect to the parameter $\lambda_q$,
the eigenvalues and eigenvectors of $\tO_q$ are given by
\begin{equation}
   \to_j^q = \omega_j^q + \bv_j^{q\dagger}\Lambda_q\bv_j^q
\end{equation}
and
\begin{equation}
  \tv_j^q = N_{qj}\left(\bv_j^q-\sum_{k\neq j}{\bv_k^{q\dagger}\Lambda_q\bv_j^q
                        \over \omega_k^q-\omega_j^q}\bv_k^q
                  \right) ,
   \label{1storder}
\end{equation}
where $N_{qj}$ is the normalization constant fixed below. For these formulae
to be acceptable, the parameters $\lambda_q$ must satisfy the restriction 
\begin{equation}
    \lambda_q^2 \ll (\omega_j^q-\omega_k^q)^2
    \label{restriction}
\end{equation}
for any pairs $(j,\,k)$ of generations.

Since $\bv_j^{q\dagger}\Lambda_q\bv_j^q = 0$ for all $j$, the perturbation
$\Lambda_q$ does not affect the eigenvalues at all, i.e.,
\begin{equation}
   \to_j = \omega_j.
\end{equation} 
Therefore, the ratios of the quark masses $m_j^q = M_q\to_j^q$, where
$M_q = \sum_j m_j^q$, are related to those of the eigenvalues
$\omega_j^q$ in Eq.~(\ref{noneigen}) by
\begin{equation}
  {m_j^q \over m_k^q} = {\omega_j^q \over \omega_k^q}.
   \quad (j,\,k = 1,\,2,\,3)
   \label{ratio}
\end{equation}
Solving Eqs.~(\ref{noneigen}) and (\ref{ratio}) in terms of the parameters
$\rho_q$ and $\th_q$, we obtain
\begin{equation}
    \rho_q = {m_3^q - m_2^q \over m_3^q + m_2^q + m_1^q}
\end{equation}
and
\begin{equation}
    \cos\th_q = \i3\left(1 + 2{m_2^q - m_1^q \over m_3^q - m_2^q}\right).
\end{equation}

To examine the behaviour of the eigenvalues and eigenvectors in response to
small deviations from the democratic limit, it is useful to introduce
parameters $\delta_q$ defiend as
\begin{equation}
    \delta_q^2 = {m_2^q - m_1^q \over m_3^q - m_2^q}.
    \label{defdelta}
\end{equation}
With these parameters, we find the expressions
\begin{equation}
   \cos{\th_q \over 2} = \sqrt{2 \over 3}
                         \left(1 + {1 \over 2}\delta_q^2\right)^\half,
   \quad
   \sin{\th_q \over 2} = \sqrt{1 \over 3}
                         \left(1 - \delta_q^2\right)^\half.
\label{halftheta} 
\end{equation}
Therefore the eigenvectors in Eq.~(\ref{noneigen}) are determined by the
parameters $\delta_q$ and the unknown phases $\phi_q$. \par

Using the experimental fact $m_1^q \ll m_3^q$, we estimate
\begin{equation}
    \rho_q = \left(1 + 2\delta_q^2 + {3m_1^q \over m_3^q - m_2^q}\right)^{-1}
          \simeq 1 - 2\delta_q^2,
\end{equation}
which results approximately in the hierarchical mass spectrum
\begin{equation}
       m_1^q : m_2^q : m_3^q
     = \omega_1^q : \omega_2^q : \omega_3^q 
     = \delta_q^4 : \delta_q^2 : 1
\end{equation}
for each $q$-sector. In short, the hierarchical order estimations
\begin{equation}
      m_j^q \propto \delta_q^{6-2j}\quad (q = u, d)
     \label{hierarchy}
\end{equation}
approximately hold. Note here that the best fittings of the quark masses
and the CKM matrix elements in \S 4 give the estimates 
$\delta_u^2 \approx 10^{-3}$ and $\delta_d^2 \approx 10^{-2}$
(see Eq.~(\ref{numerical1})).
Taking these facts into account beforehand, we have introduced the parameter
$\delta_q^2$ rather than $\delta_q$ to describe the deviation of mass
spectrum from the democratic limit in Eq.~(\ref{defdelta}). This definition
turns out to be crucial to obtain a simple representation of the CKM matrix
in terms of parameters describing small deviations from the democratic limit
and to extract sum rules from it.
\par

\section{Sum rules for absolute values of CKM matrix elements}

\qquad
For the restriction in Eq.~(\ref{restriction}) to be fulfilled,
the parameter $\lambda_q$ must be subject to the condition
$\lambda_q^2\ll (\omega_2^q-\omega_1^q)^2\approx\omega_2^{q2}\simeq\delta_q^4$.
Off-diagonal matrix elements of the perturbation $\Lambda_q$ are calculated
to be
\begin{equation}
  \bv_1^{q\dagger}\Lambda_q\bv_2^q =  \sqr2\lambda_q\cos{\th_q\over 2},\ \ 
  \bv_2^{q\dagger}\Lambda_q\bv_3^q = 0,\ \ 
  \bv_3^{q\dagger}\Lambda_q\bv_1^q = -\sqr2\lambda_q\sin{\th_q\over 2}.
\end{equation}
Using these results, we obtain the eigenvectors of the matrix $\tO_q$ in
the forms
\begin{equation}\left\{
 \begin{array}{l}
  \noalign{\vskip 0.15cm}
  \tv_1^q = N_{q1}(\bv_1^q - c_q\bv_2^q + s_q\bv_3^q),\\
  \noalign{\vskip 0.15cm}
  \tv_2^q = N_{q2}(\bv_2^q + c_q\bv_1^q),\\
  \noalign{\vskip 0.15cm}
  \tv_3^q = N_{q3}(\bv_3^q - s_q\bv_1^q),
 \end{array}
 \right.
 \label{perturbedeigen}
\end{equation}
where
\begin{equation}
  c_q = {\bv_2^{q\dagger}\Lambda_q\bv_1^q \over \omega_2^q-\omega_1^q}
      \simeq \sqr2{\lambda_q \over \omega_2^q}\cos{\th_q \over 2},
  \label{cqdefine}
\end{equation}
\begin{equation}
  s_q = {\bv_1^{q\dagger}\Lambda_q\bv_3^q \over \omega_1^q-\omega_3^q}
      \simeq \sqr2{\lambda_q \over \omega_3^q}\sin{\th_q \over 2}
      \simeq {1\over\sqr2}{m_2^q \over m_3^q}c_q
  \label{sqdefine}
\end{equation}
and
\begin{equation}
  N_{q1,2} \simeq 1 -\half c_q^2 \equiv N_q,\quad N_{q3} \simeq 1.
  \label{normalization}
\end{equation}
Here the terms $s_q^2$ are neglected in comparison with $c_q^2$.\par

In this scheme, with Hermitian mass matrices, the transformation matrix
connecting the chiral quark fields in interaction and mass eigenmodes are
constructed to be $U^q=(\tv_1^q,\,\tv_2^q,\,\tv_3^q)$. Therefore the CKM matrix
$V \equiv U^{u\dagger}U^d = (\tv_j^{u\dagger}\tv_k^d)$ is calculated
as follows:
\begin{equation}
\!\!\!\!
V\!\!=\!\!\left(
            \begin{array}{lll}
             \vspace{0.2cm}
             \begin{array}{l}
              \!\!\!\!N_uN_d(
              1 + c_uc_d\bv_2^{u\dagger}\bv_2^d\\
              \qquad - c_us_d\bv_2^{u\dagger}\bv_3^d
                     - s_uc_d\bv_3^{u\dagger}\bv_2^d)
             \end{array}
              &
             \begin{array}{l}
              \!\!\!\!N_uN_d(c_d - c_u\bv_2^{u\dagger}\bv_2^d\\
              \qquad\ +\, s_u\bv_3^{u\dagger}\bv_2^d)
             \end{array}
              &
             \begin{array}{l}
              \!\!\!\!\!\!N_u(s_u - s_d\\
              \qquad - c_u\bv_2^{u\dagger}\bv_3^d)
             \end{array}
              \\
            \noalign{\vskip 0.2cm}
                  N_uN_d(c_u - c_d\bv_2^{u\dagger}\bv_2^d
                         + s_d\bv_2^{u\dagger}\bv_3^d) &
              \!\!N_uN_d(c_uc_d + \bv_2^{u\dagger}\bv_2^d) &
              \!\!\!\!N_u(\bv_2^{u\dagger}\bv_3^d - c_u s_d)\\
            \noalign{\vskip 0.4cm}
                  N_d(s_d - s_u - c_d\bv_3^{u\dagger}\bv_2^d) &
              \!\!N_d(\bv_3^{u\dagger}\bv_2^d - s_u c_d)&
              \!\!\!\!\bv_3^{u\dagger}\bv_3^d + s_u s_d
            \end{array}
          \!\!\!\!\right)
\label{CKM}
\end{equation}
which is expressed in terms of the parameters $c_q$ and $s_q$ and the
unperturbed matrix elements $(\bv_j^{u\dagger}\bv_k^d)$ for $j,\,k = 2,\,3$.
Note that $s_q$ is given by $c_q$ and that the unperturbed matrix elements
calculated from the eigenvectors in Eq.~(\ref{noneigen}) are determined as
functions of two angles $\th_q (q = u,\,d)$ and one phase $\phi=\phi_u-\phi_d$.
Therefore the CKM matrix obtained in Eq.~(\ref{CKM}) depends on five
parameters $c_q$, $\th_q$ and $\phi$. \par

At this stage it is necessary to clarify how the unperturbed matrix elements
$(\bv_j^{u\dagger}\bv_k^d)$ depend on the small parameters representing
the deviations from the democratic limit. Retaining the linear and quadratic
terms of $\delta_q$ and $\phi$, we find
\begin{equation}
 \!\!\!\!\!\!\!\left\{
  \begin{array}{l}
    \bv_2^{u\dagger}\bv_2^d
     = \sin\displaystyle{\th_u\over 2}\sin\displaystyle{\th_d\over 2}
     + e^{i\phi}\cos\displaystyle{\th_u\over 2}\cos\displaystyle{\th_d\over 2}
     \simeq 1 -\displaystyle{1 \over 3}\phi^2 +\displaystyle{2 \over 3}i\phi,\\
    \noalign{\vskip 0.3cm}
    \bv_2^{u\dagger}\bv_3^d
     = \sin\displaystyle{\th_u\over 2}\cos\displaystyle{\th_d\over 2}
     - e^{i\phi}\cos\displaystyle{\th_u\over 2}\sin\displaystyle{\th_d\over 2}
    \simeq \displaystyle{\sqrt{2}\over 3}
       \left[\displaystyle{1 \over 2}\phi^2
        - \displaystyle{3 \over 4}(\delta_u^2 - \delta_d^2)-i\phi\right],\\
    \noalign{\vskip 0.3cm}
    \bv_3^{u\dagger}\bv_2^d
     = \cos\displaystyle{\th_u\over 2}\sin\displaystyle{\th_d\over 2}
     - e^{i\phi}\sin\displaystyle{\th_u\over 2}\cos\displaystyle{\th_d\over 2}
    \simeq \displaystyle{\sqrt{2}\over 3}
       \left[\displaystyle{1 \over 2}\phi^2
        + \displaystyle{3 \over 4}(\delta_u^2 - \delta_d^2)-i\phi\right],\\
    \noalign{\vskip 0.3cm}
    \bv_3^{u\dagger}\bv_3^d
     = \cos\displaystyle{\th_u\over 2}\cos\displaystyle{\th_d\over 2}
     + e^{i\phi}\sin\displaystyle{\th_u\over 2}\sin\displaystyle{\th_d\over 2}
    \simeq 1 - \displaystyle{1 \over 6}\phi^2 + \displaystyle{1 \over 3}i\phi.
  \end{array}
  \right.
  \label{melement}
\end{equation}
Substituting these estimations and the approximation for the normalization
constant in Eq.~(\ref{normalization}) into Eq.~(\ref{CKM}), we finally obtain
\begin{equation}
   \!\!\!\!
    \begin{array}{l}
     \noalign{\vskip -0.15cm}
     V \simeq\\
      \left(
            \begin{array}{ccc}
             \!\!\!1 - \displaystyle{1\over 2}(c_u - c_d)^2        &
            \!\!\!c_d - c_u - i\displaystyle{2\over 3}c_u\phi &
              s_u - s_d + i\displaystyle{\sqrt{2}\over 3}c_u\phi\\
            \noalign{\vskip 0.4cm}
             \!\!
             \!\!c_u - c_d - i\displaystyle{2\over 3}c_d\phi    &
            \!\!1 - \displaystyle{1\over 2}(c_u - c_d)^2
             - \displaystyle{1\over 3}\phi^2 + \displaystyle{2 \over 3}i\phi &
              \displaystyle{\sqrt{2}\over 3}
              [\displaystyle{1\over 2}\phi^2
               - \displaystyle{3 \over 4}(\delta_u^2 - \delta_d^2) - i\phi]\\
            \noalign{\vskip 0.4cm}
             s_d - s_u + i\displaystyle{\sqrt{2}\over 3}c_d\phi  &
            \!\!\displaystyle{\sqrt{2}\over 3}
              [\displaystyle{1\over 2}\phi^2
               + \displaystyle{3 \over 4}(\delta_u^2 - \delta_d^2) - i\phi] &
              \!\!1 -\displaystyle{1 \over 6}\phi^2
                    + \displaystyle{1\over 3}i\phi\\
            \end{array}\!\!
          \right)
        \end{array}
        \label{paraCKM}
\end{equation}
for the CKM matrix. Evidently this CKM matrix possesses the hierarchical
structure. From the diagonal elements to the most off-diagonal elements, 
the magnitudes of the dominant terms decrease in the
ratio $1 : |\phi| : |c_q\phi|$, where $|\phi|$ and $|c_q|$ are estimated
to be of order of $10^{-1}$ in \S 4.
It is straightforward to verify that this matrix satisfies the condition for
unitarity $V^\dagger V \simeq I$ in the present approximation.\par

For comparison with available experimental results, it is necessary to
calculate the absolute values of the matrix elements.
Note that, although $V_{cb}$ and $V_{ts}$ in Eq.~(\ref{paraCKM})
depend on $\delta_q^2$, the dependence becomes higher order and can be
ignored in $|V_{cb}|$ and $|V_{ts}|$. Consequently,
the absolute values of all the CKM matrix elements are expressed solely
in terms of the three small parameters $c_u$, $c_d$ and $\phi$.
This means that there are six independent sum rules among the absolute
values of the nine CKM matrix elements.\par

Neglecting terms higher than second order in $c_u$, $c_d$ and $\phi$,
we readily find
\begin{equation}
   2|V_{ud}| + |V_{us}|^2 \simeq 2
   \label{sumrule1}
\end{equation}
and
\begin{equation}
   |V_{us}| \simeq |V_{cd}|
   \label{sumrule2}
\end{equation}
for the first and second generations, and
\begin{equation}
     2|V_{tb}| + |V_{cb}|^2 \simeq 2
   \label{sumrule3}
\end{equation}
and
\begin{equation}
     |V_{cb}| \simeq |V_{ts}|
   \label{sumrule4}
\end{equation}
for the second and third generations.
Although $|V_{ud}|$, $|V_{us}|$ and $|V_{cd}|$ are functions of $|c_u-c_d|$,
and $|V_{tb}|$, $|V_{cb}|$ and $|V_{ts}|$ are functions of $\phi$, the formulas
in Eqs.~(\ref{sumrule1}) and (\ref{sumrule2}) are of exactly the same form
as those in Eqs.~(\ref{sumrule3}) and (\ref{sumrule4}). 
As the fifth sum rule we obtain
\begin{equation}
   2|V_{ud}| - 2|V_{cs}| \simeq |V_{cb}|^2
   \label{sumrule5}
\end{equation}
for the matrix elements of three generations. To obtain the sixth sum rule,
it is necessary to fix the sign of $c_d - c_u$. Assuming that $c_d - c_u < 0$
(see Eq.~(\ref{numerical2})) and using the expression in Eq.~(\ref{sqdefine}),
we find the hybrid sum rule
\begin{equation}
  \begin{array}{l}
   \displaystyle{
        \left({m_c \over m_t} + {m_s \over m_b}\right)|V_{us}|^2|V_{cb}|^2 -
        \left({m_c \over m_t} - {m_s \over m_b}\right)(|V_{td}|^2 - |V_{ub}|^2)
                }\\
  \noalign{\vskip 0.2cm}
   \simeq \displaystyle{
         \left[4(|V_{td}|^2 + |V_{ub}|^2)|V_{us}|^2|V_{cb}|^2
         - 2|V_{us}|^4|V_{cb}|^4
         - 2(|V_{td}|^2 - |V_{ub}|^2)^2\right]^\half|V_{cb}|}
  \end{array}
  \label{sumrule6}
\end{equation}
for the quark masses and the CKM matrix elements of three generations.

In the lowest order calculation, the rephasing invariant Jarlskog parameter
is calculated to be
\begin{equation}
  J \simeq {\sqrt{2} \over 3}(c_u - c_d)(s_u - s_d)\phi,
  \label{Jarlskog}
\end{equation}
the magnitude of which is obtained by
\begin{equation}
  \begin{array}{lll}
  |J| &\simeq& \displaystyle{{1 \over 2\sqrt{2}}
       \left[\left({m_c \over m_t} + {m_s \over m_b}\right)|V_{us}|^2|V_{cb}|
           - \left({m_c \over m_t} - {m_s \over m_b}\right)
                   {|V_{td}|^2 - |V_{ub}|^2 \over |V_{cb}|}\right]}\\
  \noalign{\vskip 0.3cm}
      &\approx&  \displaystyle{{1 \over 2\sqrt{2}}
             \left({m_c \over m_t} + {m_s \over m_b}\right)|V_{us}|^2|V_{cb}|}.
  \end{array}
  \label{magnitudeJarlskog}
\end{equation}

To derive the CKM matrix in the Wolfenstein parameterization~\cite{rf:20},
which is convenient to analyze the unitarity triangle, it is necessary
to apply a phase transformation on $V$ in Eq.~(\ref{paraCKM}) as follows: 
\begin{equation}
  \left( \begin{array}{ccc}
           -1  &   0   &   0\\
           0   &   1   &   0\\
           0   &   0   & - \displaystyle{V_{ts}^\ast \over \mid V_{ts}\mid}\\
         \end{array}
  \right)
  V
  \left( \begin{array}{ccc}
           -1  &   0   &   0\\
           0   &   1   &   0\\
           0   &   0   & \displaystyle{V_{cb}^\ast \over \mid V_{cb}\mid}\\
         \end{array}
  \right).
\end{equation}
Then we obtain the expressions
\begin{equation}
 \begin{array}{lll}
  \rho &\simeq& \displaystyle{{\Re(V_{ub}V_{cb}^\ast)\over(c_d - c_u)
                              \mid V_{cb}\mid^2}}\\
  \noalign{\vskip 0.2cm}
       &\simeq& \displaystyle{{1 \over (c_u - c_d)\mid V_{cb}\mid^2}}
                \left\{{2 \over 9}c_u\phi^2 + {\sqrt{2} \over 3}(s_d - s_u)
                [\half\phi^2 - {3 \over 4}(\delta_u^2 - \delta_d^2)]\right\},\\
  \noalign{\vskip 0.4cm}
  \eta &\simeq& \displaystyle{{\Im(V_{ub}V_{cb}^\ast)\over(c_u- c_d)
                              \mid V_{cb}\mid^2}}
        \simeq {\sqrt{2} \over 3}\displaystyle{{1 \over \mid V_{cb}\mid^2}}
               {s_d - s_u \over c_d - c_u}\phi
  \end{array}
  \label{Wolfenstein}
\end{equation}
for the Wolfenstein parameters.\par

\section{Numerical results}

\qquad
We have extracted the six (five plus one) sum rules and the expressions for
the Jarlskog and Wolfenstein parameters from the quasi-democratic mass
matrices ${\cal M}_q = M_q\tO_q$ using the lowest order approximation with
respect to the small deviations from the perfect democratic limit.
Now it is possible to set the analytical results to the experimental data
on the absolute values of the CKM matrix elements. \par

At present the world averages of the absolute values of the CKM matrix
elements have been estimated by the Particle Data Group
as follows~\cite{rf:21}:
\begin{equation}
  \left(
    \begin{array}{lll}
      0.9745\sim0.9757 & 0.219\sim0.224   & 0.002\sim0.005 \\
     \noalign{\vskip 0.15cm}
      0.218\sim0.224   & 0.9736\sim0.9750 & 0.036\sim0.046 \\
     \noalign{\vskip 0.15cm}
      0.004\sim0.014   & 0.034\sim0.046   & 0.9989\sim0.9993
    \end{array}
  \right).
  \label{world}
\end{equation}
For the mixing between $c$ and $b$ quarks the new value,
$|V_{cb}| = 0.039\pm 0.002$, has been reported~\cite{rf:22,rf:23}.  \par

Evidently the simple relations in Eqs.~(\ref{sumrule2}) and (\ref{sumrule4})
are consistent with the experimental data. Substitution of the central values
of the matrix elements into Eqs.~(\ref{sumrule1}) and (\ref{sumrule3}) leads
to $2\times0.9751 + 0.221^2 = 1.9993$ for the left hand side (lhs)
of Eq.~(\ref{sumrule1}) and $2\times0.9991 + 0.039^2 = 1.9997$ for
the lhs of Eq.(\ref{sumrule3}), respectively. As for Eq.(\ref{sumrule5}),
we obtain $2\times0.9751 - 2\times0.9743 = 0.0016$ for the lhs, which is close
to $(0.039)^2 = 0.0015$ for the right-hand side. Note that these five sum
rules are manifestation of the hierarchical structure of the CKM matrix.

Let us make a numerical estimate for a set of parameters so as to
reproduce the experimental values of both the quark masses and the CKM matrix
elements. Using a 2-loop renormalization group calculation, Fusaoka and Koide
obtained the values of quark masses at 1 GeV as~\cite{rf:24}
\begin{equation}
 \begin{array}{lll}
     m_u = 4.90\pm0.53\ {\rm MeV}, & m_c = 1467\pm28^{+5}_{-2}\ {\rm MeV},
   & m_t = 339\pm24^{+12}_{-11}\ {\rm GeV}, \\
    \noalign{\vskip 0.3cm}
     m_d = 9.76\pm0.63\ {\rm MeV}, & m_s = 187\pm16\ {\rm MeV},
   & m_b = 6356\pm80^{+214}_{-164}\ {\rm MeV}.
 \end{array}
 \label{Koide}
\end{equation}
The best fittings of the absolute values of the CKM matrix elements in
Eq.~(\ref{paraCKM}) to the world averages in Eq.~(\ref{world}) and of
the quark masses calculated by Eqs.~(\ref{noneigen}) and (\ref{ratio}) to
the values in Eq.~(\ref{Koide}) are accomplished by choosing
\begin{equation}
 \left\{
 \begin{array}{lllll}
  \rho_u &=& 0.9914 &=& 1 - 8.632\times 10^{-3},\\
  \noalign{\vskip 0.2cm}
  \rho_d &=& 0.9414 &=& 1 - 5.856\times 10^{-2},\\
  \noalign{\vskip 0.2cm}
  \cos\theta_u &=& 0.3362 &=& \displaystyle{1 \over 3} + 2.888\times 10^{-3},\\
  \noalign{\vskip 0.2cm}
  \cos\theta_d &=& 0.3525 &=& \displaystyle{1 \over 3} + 1.915\times 10^{-2},
 \end{array}
 \right.
 \label{numerical1}
\end{equation}
\begin{equation}
 \left\{
 \begin{array}{lllll}
  \lambda_u &=& 2.37\times10^{-4}\ \ 
              &( c_u = \ \ 0.0643,\ & s_u = \ \ 1.939 \times 10^{-4} )\\
  \noalign{\vskip 0.2cm}
  \lambda_d &=& - 3.64\times 10^{-3}\ 
              &( c_d = - 0.1564,\ & s_d = - 3.024 \times 10^{-3} )
 \end{array}
 \right.
 \label{numerical2}
\end{equation}
and
\begin{equation}
   \phi \equiv \phi_u - \phi_d = 7.99\times 10^{-2}.
 \label{numerical3}
\end{equation}
The smallness of $|\lambda_u|$ and $|\lambda_d|$ confirms the validity of
the lowest order perturbation approximation. From the set of parameters in
Eqs.~(\ref{numerical1}) $\sim$ (\ref{numerical3}), we reproduce
\begin{equation}
    \left(
    \begin{array}{lll}
      0.9756  &  0.2207  &  0.0040\\
     \noalign{\vskip 0.15cm}
      0.2209  &  0.9749  &  0.0377\\
     \noalign{\vskip 0.15cm}
      0.0067  &  0.0377  &  0.9993
    \end{array}
    \right)
 \label{numericalresult}
\end{equation}
for the magnitudes of the CKM matrix elements in Eq.~(\ref{paraCKM}).
This result with the data in Eq.~(\ref{Koide}) enables us to calculate
the absolute value of the Jarlskog parameter as
\begin{equation}
     |\,J\,|\, \simeq 2.67 \times 10^{-5}
\end{equation}
by Eq.~(\ref{Jarlskog}) and its order to be $|J|\approx 2.2\times 10^{-5}$
by the second relation of Eq.~(\ref{magnitudeJarlskog}). We obtain
\begin{equation}
     \rho \simeq 0.187,\quad \eta \simeq 0.386
\end{equation}
for the Wolfenstein parameters in Eq.~(\ref{Wolfenstein}). All these results
are consistent with experimental results. Finally, the left and right hand
sides of the hybrid sum rule in Eq.(\ref{sumrule6}) are estimated,
respectively, to be $3.1\times 10^{-6}$ and $2.9\times 10^{-6}$. \par

\section{Discussion}

\qquad
In this way we have developed a simple formalism for the quark masses
and the weak mixing matrix which explains all the characteristic features
of quark flavours quite well. The mass matrices are postulated to be
self-adjoint and to have a democratic skelton with small deviations
from it. It is essential to parametrize the mass matrices and
to decompose them into the unperturbative and perturbative parts
so that the mass eigenvectors are obtained in simple mathematical forms.
The mass eigenvalues $m_j^q = M_q\,\omega_j^q$ which are not sensitive to
perturbations have hierarchical structures and approximately obey the power
laws $m_j^q \propto \delta_q^{6-2j}$ with respect to the small parameters
$\delta_q^2$. The CKM martix elements also exhibit hierarchical structures
when expanded with respect to small deviations from the democratic limit.\par

It is worthwhile to observe that the hierarchical structures appear
distinctively in the mass spectra of up and down quark sectors and in
the CKM matrix elements. The up and down quark sectors have
the mass spectra characterized, respectively, by of the parameter values
$\delta_u^2 \simeq 4.33\times10^{-3}$ and
$\delta_d^2 \simeq 2.87\times10^{-2}$.
As for the strength of the perturbations, $|\lambda_u|$ is smaller
than $|\lambda_d|$ by one order. Accordingly, the up quark sector
is more democratic and hierarchical than the $d$-sector.
On the other hand, the dominant terms in the CKM matrix decrease,
from the diagonal elements to the most off-diagonal elements, in the ratio
$1\,:\,|\phi|\,:\,|c_q\,\phi|$, where $|\phi|$ and $|c_q|$ are of order
$10^{-1}$. Therefore the behaviour of the CKM matrix is less democratic
and hierarchical than that of the quark mass spectra. \par

The absolute value of the CKM matrix elements turn out to be expressible
in terms of three parameters, $c_u, c_d$ and $\phi$, in our approximation.
This remarkable fact enables us to derive six independent sum rules
among the nine matrix elements. The sum rules in
Eqs.~(\ref{sumrule1})$\sim$(\ref{sumrule5}) are rather general consequences
of the hierarchical structure of the CKM matrix.
The sixth sum rule in Eq.~(\ref{sumrule6}) and the representation
of the Jarlskog parameter in Eq.~(\ref{magnitudeJarlskog}) are unique
results of our method and consistent with the present experimental data.
The set of parameters in Eqs.~(\ref{numerical1})$\sim$(\ref{numerical3})
reproduces both the observed data of the quark mass spectra and the CKM matrix
elements to high accuracy. From these results it is not unreasonable to infer
that our quasi-democratic mass matrices reflect essential elements of
quark flavour physics.\par

As a next step of our approach, we must clarify whether or not the strong
restrictions imposed on the quasi-democratic mass matrices such as their
self-adjointness and the restriction on phases in Eq.~(\ref{phase}) are
indispensable. At this stage, when the origin of the quasi-democratic mass
matrices is not known, it is merely a matter of convenience to assume
the mass matrices to be self-adjoint. It is necessary to investigate various
models with left-right symmetric and asymmetric quasi-democratic mass matrices.
As for the phase restriction, we must examine many other models with mass
matrices containing different entries of CP violating phases. Such attempts
at model building and comparisons of the models with detailed experimental
data bring us closer to the true nature of quark flavour physics.
We will study one possible origin of the quasi-democratic mass matrices
in Eq.~(\ref{quasi-democratic}) in a future publication.\par


\begin{thebibliography}{99}
\bibitem{rf:1}
  M.~Kobayashi and K.~Maskawa, Prog. Theor. Phys. {\bf 49} (1973), 652.
\bibitem{rf:2}
  F.~J.~Gilman, K.~Kleinknecht and B.~Renk, Phys. Rev. {\bf D54} (1996), 94,
  and references therein.
\bibitem{rf:3}
  H.~Fritzsch, Phys. Lett. {\bf B73} (1978), 317; {\bf B85} (1979), 81; 
               {\bf B353} (1995), 114.
\bibitem{rf:4}
  M.~Shin, Phys. Lett. {\bf B145} (1984), 285. 
\bibitem{rf:5}
  B.~Stech, Phys. Lett. {\bf 130B} (1983), 189.
\bibitem{rf:6}
  M.~Gronau, R.~Johnson and J.~Schechter, Phys. Rev. Lett. {\bf 54}
  (1985), 2176.
\bibitem{rf:7}
  H.~Harari, H.~Haut and J.~Weyers, Phys. Lett. {\bf 78B} (1978), 459.
\bibitem{rf:8}
  T.~Goldman and G.~J.~Stephenson,~Jr., Phys. Rev. {\bf D24} (1981), 236.
\bibitem{rf:9}
  Y.~Koide, Phys. Rev. Lett. {\bf 47} (1981), 1241; Phys. Rev. {\bf D28}
  (1983), 252; {\bf 39} (1989), 1391.
\bibitem{rf:10}
  P.~Kaus and S.~Meshkov, Mod. Phys. Lett. {\bf A3} (1988), 1251;
  Phys. Rev. {\bf D42}  (1990), 1863.
\bibitem{rf:11}
  L.~Lavoura, Phys. Lett. {\bf B228}  (1989), 245.
\bibitem{rf:12}
  M.~Tanimoto, Phys. Rev. {\bf D41} (1990), 1586.
\bibitem{rf:13}
  H.~Fritzsch and J. Plankl, Phys. Lett. {\bf B237} (1990), 451.
\bibitem{rf:14}
  Y. Nambu, in {\it Proceedings of the International Workshop on Electroweak
  Symmetry Breaking}, Hiroshima (World Scientific, Singapore, 1992), p.1.
\bibitem{rf:15}
  G.~C.~Branco, J.~I.~Silva-Marcos and M.~N.~Rebelo, Phys. Lett. {\bf B237}
  (1990), 446.
\bibitem{rf:16}
  J.~Kalinowski and M.~Olechowski, Phys. Lett. {\bf B251} (1990), 584.
\bibitem{rf:17}
  G.~C.~Branco and J.~I.~Silva-Marcos, Phys. Lett. {\bf B359} (1995), 166.\\
  G.~C.~Branco, D.~Emmanuel-Costa and J.~I.~Silva-Marcos, Phys. Rev.
  {\bf D56} (1997), 107.
\bibitem{rf:18}
  P.~M.~Fishbane and P.~Kaus, Phys. Rev. {\bf D49} (1994), 3612, 4780;
  Z. Phys. {\bf C75} (1997), 1.
\bibitem{rf:19}
  T.~Teshima and T.~Sakai, Prog. Theor. Phys. {\bf 97} (1997), 653.
\bibitem{rf:20}
  L.~Wolfenstein, Phys. Rev. Lett. {\bf 51} (1983), 1945.
\bibitem{rf:21}
  Particle Data Group, R.~M.~Barnett  et al., Phys. Rev. {\bf D54}
  (1996), 1.
\bibitem{rf:22}
  M.~Neubert, Int. J. Mod. Phys. {\bf A11} (1996), 4173.
\bibitem{rf:23}
  R.~Forty, Talk given at {\it the Second International Conference on
  B Physics and CP Violation}, Honolulu, Hawaii, March 24-27, 1997.
\bibitem{rf:24}
  H. Fusaoka and Y.~Koide, hep-ph9706211.
\end{thebibliography}
\end{document}